\documentstyle[preprint, 12pt, aps, epsfig]{revtex} 
\topmargin 
-10pt
\oddsidemargin 
-10pt
\newcommand{\be}{\begin{equation}}
\newcommand{\ee}{\end{equation}}
\newcommand{\ba}{\begin{eqnarray}}
\newcommand{\ea}{\end{eqnarray}}

\newcommand{\di}{\displaystyle}
\newcommand{\bml}{\begin{mathletters}}
\newcommand{\eml}{\end{mathletters}}
\def\O{{\cal O}}
\tightenlines

\begin{document}

\begin{center} 
{\Large {\bf Lattice 
Kinetics of Diffusion-Limited Coalescence and 
Annihilation
with 
Sources }}\\[1cm]
E. Abad\footnote[1]{E-mail: 
eabad@ulb.ac.be }\\
{\it Centre for Nonlinear Phenomena and 
Complex Systems\\
Universit\'e Libre de Bruxelles\\
Campus 
Plaine C.P. 231\\
B-1050 Bruxelles\\
} 
T. Masser and D. 
ben-Avraham\footnote[2]{E-mail: benavraham@clarkson.edu 
}\\
{\it
Physics Department and Clarkson Institute for 
Statistical Physics\\
Clarkson University, Potsdam, NY 
13699-5820\\
}
\today\\
\end{center}

{\bf Abstract:}
We study the 1D 
kinetics of diffusion-limited coalescence and annihilation 
with back reactions and different kinds
of particle input.  
By considering the changes in occupation and 
parity of a 
given interval, 
we derive sets of hierarchical equations 
from which exact expressions 
for the lattice coverage and 
the particle concentration can be obtained. 
We compare the 
mean-field approximation and the continuum 
approximation to 
the exact solutions and we  discuss their regime of 
validity. \\[.5cm]

PACS 
number(s): 02.50.Ey, 05.50.+q, 05.70.Ln, 
82.40.-g\\

\section{Introduction}

A standard way of studying the kinetics of 
reaction-diffusion systems
involves derivation of an infinite set 
of moment equations for the state
probability. The solution 
of such equations usually poses great 
mathematical
difficulties; however, the simple topology of the one-dimensional 
lattice frequently
allows derivation of exact solutions \cite{Priv0,Schu,Marr,Hen1,Alc,Cadi}.  
These results may be used to test the
validity of various 
approximations. Neglecting spatial fluctuations in 
concentration
and occupation number space leads to classical 
macroscopic rate equations. This type
of approximation, which 
implicitly assumes that each particle interacts with the
system 
as a whole through a  {\it mean field} (MF), should improve 
as the number of
interacting neighbors grows, i.e., with 
increasing dimensionality of the lattice.
Although this 
approach may seem too rough for
one-dimensional systems, it 
works well for early time evolution and in other
situations 
where the distribution of particles is nearly free of 
correlations.     

Exactly solvable reaction-diffusion models 
consist largely of single species
reactions in one dimension, 
e.g., variations of the coalescence process,
$A+A\rightarrow 
A+S$ \cite{Avr,Avr2,Doer,Kreb,Sim,Mobi} and the annihilation 
process
$A+A\rightarrow S+S$ 
\cite{Kreb,Sim,Torn,Lush,Racz,Spou,Fam,Redn,Priv,Gryn,Hen2}, 
where $A$ and $S$ denote empty and
occupied sites, 
respectively. These simple reactions display a wide range of 
behavior
characteristic of non-equilibrium kinetics, such as self 
organization,
pattern formation, and kinetic phase 
transitions. Interval methods have provided
many exact 
solutions for one-dimensional coalescence and annihilation 
models. The
method of empty intervals, applicable to 
coalescence models, requires solution of an
infinite hierarchy of 
differential  difference equations for the probabilities 
$E_n$
of finding $n$ consecutive lattice sites simultaneously 
empty \cite{Avr,Avr2,Doer,Mobi,Lin,Pes,Ali}. For annihilation 
models, the method of parity intervals
similarly requires determination 
of $G_n$, the probability of $n$ consecutive
lattice sites 
containing an even number of particles \cite{Mass,Mass2,Habib}.

The 
infinite hierarchy of equations for the interval 
probabilities can be used as a 
starting point for the derivation of 
MF  and continuum approximations.  Most results
obtained 
from interval methods rely on a continuum approximation of 
the spatially
discrete equations; the resulting PDEs allow 
for straightforward solution. However,
the pertinent set of 
differential difference equations for the 
interval probabilities admits analytical solutions
\cite{Abad}, providing 
exact results which may differ from the  continuum limit. 
In
general, we expect that this discrepancy becomes important 
for high concentrations
and short time scales (compared to 
the typical reaction time), since on-lattice
local 
concentration perturbations travel at
finite  speed, in contrast to 
propagation in a spatial continuum. 

In this work, we use the interval methods described above to study the 
on-lattice kinetics of coagulation and annihilation reactions 
including particle creation steps. Explicit expressions for the 
lattice coverage and the concentration are derived and compared to 
those yielded by
the MF and the continuum approximation. The behavior of coalescence 
and annihilation changes qualitatively if 
the reaction schemes incorporate 
particle sources in the form of input (external
source) or back  reactions (internal 
source). In the absence of sources, the
concentration displays an
anomalous $t^{-1/2}$-decay to an empty, steady state 
\cite{Redn,Priv}.  Particle
sources give rise to steady 
states with nonzero concentration
and change the transient 
behavior \cite{Doer,Racz}.  In most cases, the concentration may
be thought of 
as an order parameter in a  ``phase transition'' between 
the empty
and the active steady state, which is controlled by  
the effective rate of particle
creation, $h$. In analogy to 
the theory of critical phenomena, one can regard $h=0$
as a 
transition point and characterize the steady state and the 
relaxation time of
the system near the critical point 
through a set of static and  dynamical exponents
\cite{Smol}. We shall see that our method allows the  
characterization of the system not only near the phase transition, 
but also far beyond criticality.    

This work is organized as follows. 
In section \ref{coamod}, we define the general 
form of the 
coalescence model and solve for the cases in which the back 
reaction or
a homogeneous particle input is present.  The 
validity of MF-like rate equations
and the continuum 
approximation is analyzed vis a vis the exact solution. 
In 
section \ref{annmod}, we do the same for the annihilation model 
with different
kinds of particle input.   Similarities with 
Glauber-type spin models and the
relevance of these models 
to physical systems are briefly
discussed.  We also treat 
an annihilation model with symmetric birth
\cite{Sudb}, both  
with immobile and diffusing reactants.  We conclude in 
section \ref{sumout} with a summary of our results 
and possible extensions of this work. 

\section{Coalescence 
Models}
\label{coamod}

All reactions in sections \ref{coamod} and 
\ref{annmod} are defined on an infinite 
1D lattice with 
spacing $a$.   The basic coalescence model involves particles 
moving
randomly and asynchronously to nearest neighbor sites 
with hopping rate $2D/a^2$; in
an extended model, particles 
give birth, i.e., a particle at a given site 
generates
offspring at an empty adjacent site at rate $v/a$.  One can 
also include a
homogeneous source: particles are injected into 
the lattice at rate $Ra$.
In all cases, a particle disappears 
whenever it lands on another. Let $E_n(t)$
be the 
probability that a randomly chosen segment of $n$ consecutive  sites 
contains
no particles. By noting the changes in $E_n$ 
during a small time interval,
$\Delta t$, and taking the 
continuous time limit 
$\Delta t \to 0$, we derive the Master 
equation \cite{Avr}:
\be
\label{mast_coal}
\frac{d 
E_{n}}{dt}=\frac{2D}{a^2}(E_{n+1}-2E_{n}+E_{n-1})                       
+\frac{v}{a}(E_{n+1}-E_{n})-RnaE_{n}.
\ee
The first 
term on the rhs represents the effect of the net particle flow 
into and 
out of an empty interval, whereas the second and 
third terms describe the effect of
the back reaction 
(cooperative particle birth) and a homogeneous particle 
input,
respectively. By comparing (\ref{mast_coal}) for $n=1$ to the 
changes
in $E_1$ during a time interval 
$\Delta t$, one 
arrives at the boundary condition $E_{0}(t)=1$. 
Also, for 
non-empty lattices, $E_{\infty}(t)=0$.
The case $D>0, v=R=0$ 
has already been solved in a previous paper \cite{Abad}. 
In  that case, the reaction displays universal, anomalously 
slow $t^{-1/2}$-decay to
an empty state,  as opposed to the 
MF $t^{-1}$ asymptotic decay. This result holds
both 
on-lattice and in the continuum limit \cite{Abad}, and is 
reminiscent of
persistent transient fluctuations induced by the 
dynamic self-ordering of the system
\cite{Avr}.
   
In the sequel, we focus on the cases $D>0, v>0, R=0$ (reversible 
coalescence
$A+A\rightleftharpoons A+S$, subsection \ref{revc} ) and $D>0, v=0, R>0$ 
(coalescence with input, subsection \ref{coainp} ).
We use the initial 
condition
\be
\label{initco} 
E_{n}(0)=(1-\rho_0)^n.
\ee 
This 
corresponds to a random homogeneous distribution of particles 
characterized by 
a global coverage $\rho_0$ and a concentration 
(number of particles per unit length)
$c_0=\rho_0/a$. Our 
primary goal is to compute the time evolution of the 
global
coverage $\rho(t)=1-E_1(t)$ and the associated 
concentration $c(t)=\rho(t)/a$.

\subsection{Reversible Coalescence}

\label{revc}
 
The existence of a kinetic phase transition for this system 
in the continuum 
limit is well known \cite{Avr2}. 
Recently, Lin has shown that the phase transition is
not an 
artifact of the continuum limit, since asymptotically this 
approximation only
gives rise to quantitative corrections 
\cite{Lin2}. To do so, he solved the 
corresponding empty interval 
hierarchy by mapping it into a spin system with Glauber 
dynamics. However, it is possible to solve the hierarchy 
more
straightforwardly by means of a simple Laplace transform (LT) method, as we 
show below.  

The boundary value problem (BVP) of interest 
is \cite{Avr2}
\be
\label{revdyn}
\frac{d 
E_n}{d\tau}=E_{n+1}-2E_{n}+E_{n-1} + h\,(E_{n+1}-E_{n})\;,
\ee
where 
$\tau=2Dt/a^2$ is a dimensionless time and $h=va/2D$ is the 
relative feed 
rate; the boundary conditions read $E_0(\tau)=1$ 
and 
$E_{\infty}(t)=0$, and the random homogeneous initial 
condition is that given
by~(\ref{initco}).   First, we solve 
for the steady state $E_{n,s}$.  Setting 
${\displaystyle 
\frac{dE_n}{d\tau}=0} $
on the lhs of Eq.~(\ref{revdyn}), we 
get 

\be
(1+h)E_{n+1,s}-(2+h)E_{n,s}+E_{n-1,s}=0.
\ee
This difference equation has the 
solution
\be
E_{n,s}=(1+h)^{-n},
\ee
whence the steady state concentration is 
obtained:
\be
\label{ssconc}
c_s=\frac{\rho_s}{a}=\frac{1-E_{1,s}}{a}=\frac{v}{2D+va}.
\ee
It has been shown that this is a true 
equilibrium steady state characterized by the presence
of detailed balance and a maximum of the entropy \cite{Avr2}. 
The spatial coherence induced by the coalescence reaction 
is eventually destroyed by the back reaction. 

Eq. (\ref{revdyn}) may be solved by standard techniques, for 
example, by taking the LT with respect to $\tau$, fitting 
a power ansatz to the resulting difference 
equation, and finally inverting the LT of 
the
solution.  With the boundary data $E_0(t)=1$ and 
$E_{\infty}(t)=0$, the LT of
$E_1(t)$ is given by 

\be
\hat{E}_{1}(s)=\left( \frac{1}{s}-
\frac{1}{s+\beta} 
\right)\left(\frac{s+2+h-\sqrt{(s+2+h)^2-4(1+h)}}{2(1+h)}
\right) 
+ 
\frac{(1-\rho_0)}{s+\beta},
\ee
where $\beta = h\rho_0 - 
\rho_0^2/(1-\rho_0)$.  Using the faltung theorem for the LT 
\cite{Cars}, this can be inverted to 
yield
\begin{eqnarray}
E_{1}(\tau) & = & \frac{1}{\sqrt{1+h}}\int_{0}^{\tau} e^{-(2+h)\tau'}
I_{1}(2\sqrt{1+h}\tau')\frac{d\tau'}{\tau'}   \nonumber \\
\label{dynE1}                            
 &  & + e^{-\beta\tau}\left[ 1-\rho_0- 
\frac{1}{\sqrt{1+h}}\int_{0}^{\tau} e^{-(2+h- 
\beta)\tau'}I_{1}(2\sqrt{1+h}\tau')\frac{d\tau'}{\tau'}\right],
\end{eqnarray}
where $I_n(\cdot)$ 
is the $n$-th order modified Bessel function. If the 
lattice is
completely filled initially ($\rho_0\to 1$), the 
difference equations for
$\hat{E}_{n}(s)$ are homogeneous and 
the second term on the rhs of (\ref{dynE1})
vanishes. 

Using 
the $s\rightarrow\infty$ expansion of $\hat{E}_{1}$ 
facilitates investigation of the short time behavior:  

\be
\hat{E}_1(s)=\frac{1-\rho_0}{s}-\frac{\beta(1-\rho_0)}{s^2}+\O(s^{-3}).
\ee
By 
virtue of a Tauberian theorem \cite{Doe,Fell}, we obtain  

\be
\label{stocc}
\rho(\tau)=\rho_0 
+\left[h\rho_0(1-\rho_0)-\rho_0^2\right]\tau+\O(\tau^2),
\qquad\tau\to0 ,
\ee
for the short-time site occupancy. 

In the opposite limit, the 
expansion of $\hat{E}_{1}(s)$ about $s=0$ provides no 
additional information about the long time solution; only the 
steady state solution
is recovered.  Determining the long 
time asymptotics requires a bit more
work. We write 
$E_{1}(\tau)$ as 
\ba
E_{1}(\tau)&=&1-\rho_{s}-\frac{2}{\gamma}\int_{\tau}^{\infty} e^{-        
 \alpha\tau'}I_{1}(\gamma\tau')\frac{d\tau'}{\tau'} \nonumber \\
&&+\left[1-\rho_0-U+ \frac{2}{\gamma}\int_{\tau}^{\infty} e^{-
(\alpha-\beta)\tau'}I_{1}(\gamma\tau')\frac{d\tau'}{\tau'}\right] 
e^{-\beta\tau},
\ea
with the 
notation
\be
\alpha=2+h\quad
\gamma=2\sqrt{1+h}\quad
\rho_s=\frac{h}{1+h}\quad
U=\frac{2}{\gamma}\int_{0}^{\infty}
e^{-(\alpha-\beta)\tau'}I_{1}(\gamma\tau')\frac{d\tau'}{\tau'}.
\ee
All 
integrals in $E_{1}(\tau)$ converge, since 
$\alpha-\beta\geq\gamma$. Moreover,
$U$ is a known LT:  

\be
U=\frac{2}{\gamma^2}(\alpha-\beta-\sqrt{(\alpha-\beta)^2-\gamma^2}).
\ee
Based 
on the value of the initial coverage, $\rho_0$, there are 
three cases for $U$:
\be
U=\left\{ \begin{array}{ll}
            
1-\rho_0     &    \mbox{for $\rho_0>\rho_c,$}\\
            
\frac{1}{\sqrt{h+1}}  &    \mbox{for $\rho_0=\rho_c,$}\\
            
\frac{1}{(h+1)(1-\rho_0)} & \mbox{for $\rho_0<\rho_c,$}
           
\end{array} \right.
\ee
where $\rho_c \equiv 
1-\sqrt{1-\rho_s}$ is a critical initial coverage.
One has 
$\rho_c<\rho_s$, and $\rho_c\to\case{1}{2}\rho_s$ as $h\to0$.

For long times, the asymptotic form of $I_{1}(z)$ 
\cite{Abra} may be used to approximate the remaining integrals 
in the expression for $E_1(\tau)$.  We then have the 
following asymptotic formula for the 
coverage:
\be
\label{asform}
\rho(\tau)-\rho_s\approx\left\{ \begin{array}{ll}
-\sqrt{\frac{2}{\pi\gamma^3}}
\frac{\beta\tau^{-3/2}e^{-(\alpha-\gamma)\tau}}
{(\alpha-\beta-\gamma)(\alpha-\gamma)}    
&    \mbox{for 
$\rho_0>\rho_c,$} \\
 - \sqrt{\frac{8}{\pi\gamma^3}}\tau^{-1/2} 
e^{-(\alpha-\gamma)\tau}  
&    \mbox{for $\rho_0=\rho_c,$}\\ 

-\left(1-\rho_0-\frac{1}{(h+1)(1-\rho_0)}\right)e^{-\beta\tau} & 
\mbox{for 
$\rho_0<\rho_c.$} \end{array} \right.
\ee
Considering then 
the (dimensionless)
relaxation time $\tau_R$, defined as   

\be
\tau_R^{-1}\equiv -\lim_{\tau\to\infty}\tau^{-1} \ln | 
\rho(\tau)-\rho_s |,
\ee
we see that the system undergoes 
a second-order dynamical phase transition, from a
slowly  
relaxing phase --- with $\tau_R$ dependent on $\rho_0$ --- 
for
$\rho_0<\rho_c$, to a phase with constant $\tau_R$, for 
$\rho_0>\rho_c$.  
It can be readily shown that the first 
derivative of $\tau_R$ at $\rho_0=\rho_c$ is
continuous, 
while the second derivative displays a finite jump determined 
by the
value of $h$.

\subsubsection*{MF and continuum 
approach} 

We now examine the simplest type of MF 
approximation for reversible coalescence.
In this approach, the 
kinetics of  $\rho$ may be obtained either 
by means of simple 
heuristic arguments or directly from the empty interval 
hierarchy. Let $P_n$ be the probability of $n$ consecutive 
sites being 
simultaneously occupied. Making use of the 
identities $E_1=1-P_1$ and 
$E_2=1-2\,P_1+P_2$ in the evolution 
equation for $E_1$, we 
have
\be
\frac{dP_1}{d\tau}=-(1+h)P_2+h\,P_1.
\ee
Neglecting spatial correlations amounts to 
setting $P_2=P_1^2$ in this equation.  
Taking into account 
that $P_1$ is identical with the mean site occupation 
$\rho$, 
we get:  

\be
\frac{d\rho}{d\tau}=-(1+h)\rho^2+h\,\rho.
\ee
The physically acceptable steady state of this 
equation is given by 

\be
\rho_s^{MF}=\rho_s=\frac{h}{1+h}.
\ee
The time dependent solution matching this steady state is  

\be
\rho^{MF}=\frac{h}{\left[(h/\rho_0)-1-h 
\right]e^{-h\tau}+1+h}.
\ee
For short times, the Taylor 
expansion of the exponential 
function yields (\ref{stocc}); as expected, the MF approach is a 
good approximation at early times. In the opposite, 
long-time limit, an exponential relaxation towards the equilibrium 
steady state 
occurs:

\be
\label{declaw}
\rho^{MF}-\rho_s
\propto 
\rho_s\left[1-\frac{\rho_s}{\rho_0}\right]e^{-h\tau},\qquad\tau\to\infty.
\ee 
Thus, in the MF approach, 
$\tau_R$ is given by 
\be
\tau_R=h^{-1}.
\ee
This result is 
universal in the sense that the relaxation time does 
not
``remember''  the initial condition $\rho_0$, which is only 
contained in the
prefactor 
$1-\rho_s/\rho_0$. 

\begin{figure}[htbp]
\centerline{\psfig{figure=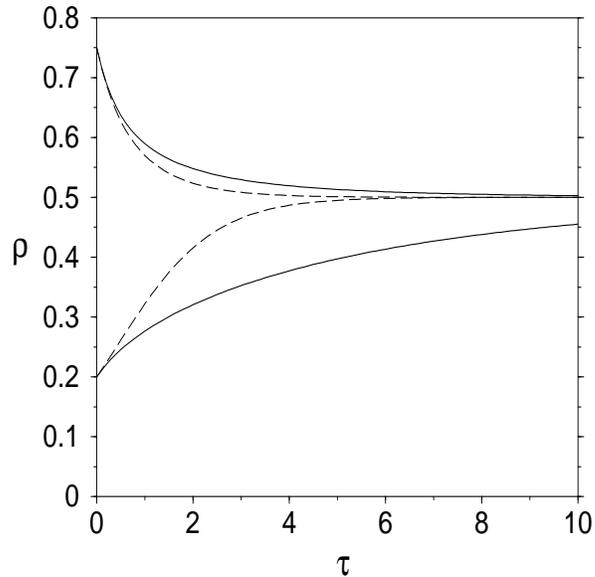,width=8cm,height=8cm}}
\caption{\label{fig1} 
Exact lattice coverage (solid lines) vs. mean field 
approximation
(dashed lines) of the reversible coalescence for 
two different initial 
conditions $\rho_0=0.75$ and 
$\rho_0=0.2$ ($h=1$). Since 
$0.2<0.29\approx \rho_c$,
the exact 
solution relaxes much slowlier into the equilibrium steady 
state in the
latter 
case}
\end{figure}

If the 
initial coverage is low enough, the relaxation towards 
equilibrium is basically given by the time taken to fill large gaps 
through particle birth from the edge sites \cite{Avr2}. 
Therefore, $\tau_R$ becomes initial-condition dependent in 
this regime, in contrast to the MF result (c.f. Fig. 
\ref{fig1}). Moreover, $\tau_R$ becomes arbitrarily large in the 
limit $\rho_0\to 0$. On the other hand, if $\rho_0>\rho_c$, 
$\tau_R$ is MF-like only for large $h$. Note, however, that 
the relaxation is not purely exponential,
since additional 
powers of $\tau$ appear in the asymptotic form  of the 
exact
solution of Eq.~(\ref{asform}). 

The exact results can also be compared with the continuum 
approximation, where one
lets  the lattice constant $a$ shrink to zero \cite{Lin2}. 
To this end, we
rewrite~(\ref{stocc}) in terms of 
concentration, using
$c_s=(1/a)\,h/(1+h)$ and $2D/v^2$ as the units 
of concentration and 
time,
respectively:
\be
c=c_0+\frac{c_0(1-c_0)}{h}t+\O(t^2), 
\ee
where $c$ and $t$ are 
dimensionless now.  On the other hand, rescaling $c$ and
$t$ as 
above and  taking the limit $h\to 0$ for fixed $R$ and $D$ in the long 
time form of the
solution, Eq.~(\ref{asform}), 
yields
\be
\label{asformc}
c=\left\{ \begin{array}{ll}
            
1+2\pi^{-1/2}[1-(1-2c_0)^{-2}]t^{-3/2}e^{-t/4}
 &  \mbox{for $c_0>1/2,$} 
\\
            1-\pi^{-1/2}t^{-1/2}e^{-t/4}  &    \mbox{for 
$c_0=1/2,$}\\ 
 1-(1-2c_0)e^{-c_0(1-c_0)t} & 
\mbox{for 
$0<c_0<1/2.$} \end{array} \right.
\ee   
An alternative 
way to perform the continuum approximation consists of 
replacing the set of difference equations (\ref{revdyn}) by the 
PDE
\be
\label{conteq}
\frac{\partial E(x,t)}{\partial t}= 
2D \frac{\partial^2 E}{\partial 
x^2}
+v \frac{\partial 
E}{\partial x},
\ee 
for a spatially continuous function 
$E(x,t)$, the boundary conditions now being
$E(0,t)=1$ and 
$E(\infty,t)=0$. This equation can be solved for $E(x,t)$
with 
a suitable exponential ansatz \cite{Avr2}; the global 
concentration 
$c(t)$ is then obtained by deriving with respect 
to space: 
\be
c(t)=-\left. \frac{\partial E(x,t)}{\partial 
x}\right|_{x=0}.
\ee
Eventually, one can rescale $c$ and 
$t$ in dimensionless units as done above 
to obtain
\be 
c(t)=1-\frac{1}{2}\mbox{erfc}\left(\frac{1}{2}\sqrt{t}\right)+
\left|\frac{1}{2}-c_0
\right| 
e^{-c_0(1-c_0)t}\,\left[\mbox{erfc}\left(\left|\frac{1}{2}-
c_0\right|\sqrt{t}\right)-2\Theta\left(\frac{1}{2}-c_0\right)\right],
\ee
where 
$\Theta(\cdot)$ is the Heaviside step function. 
As 
expected, the long time development of this expression is 
identical to 
(\ref{asformc}). In contrast, for short times one 
has
\be
c(t)=c_0+\left(\frac{1}{2\sqrt{\pi}}-2\frac{(1/2-c_0)^2}{\sqrt{\pi}}\right)
\sqrt{t}+(1/2-c_0)c_0(1-c_0)t+\O(t^{3/2}).
\ee
Thus, 
for early times there is a qualitative discrepancy with 
respect to the
on-lattice case, even as $c_0\to 0$.  The 
evolution is initially faster in a
continuum, due to the infinite 
speed of propagation of local
perturbations.

\subsection{Coalescence with input} 

\label{coainp}

As a second 
example, we consider the effect of a homogeneous external input 
$S\rightarrow A$, at rate $Ra$, on the simple coalescence 
reaction. This 
model is described by the set of equations 
\cite{Avr}: 
\be
\label{diffso}
\frac{dE_n}{d \tau}= 
E_{n-1}-(2+h\,n)\,E_n+E_{n+1},
\ee
with the boundary conditions 
$E_0=1$ and $E_\infty=0$, where we used the
dimensionless 
time $\tau=2Dt/a^2$ and relative feed rate $h=Ra^3/2D$.

We now compute the exact steady state solution by setting 
the lhs of (\ref{diffso}) 
equal to zero. The resulting set 
of difference equations can be compared to
recursion 
relations satisfied by the Bessel functions 
\cite{Abra}:
\be
J_{\nu-1}(z)-\frac{2\nu}{z}\,J_\nu(z)+J_{\nu+1}(z)=0.
\ee   

\begin{figure}[htbp]
\centerline{\psfig{figure=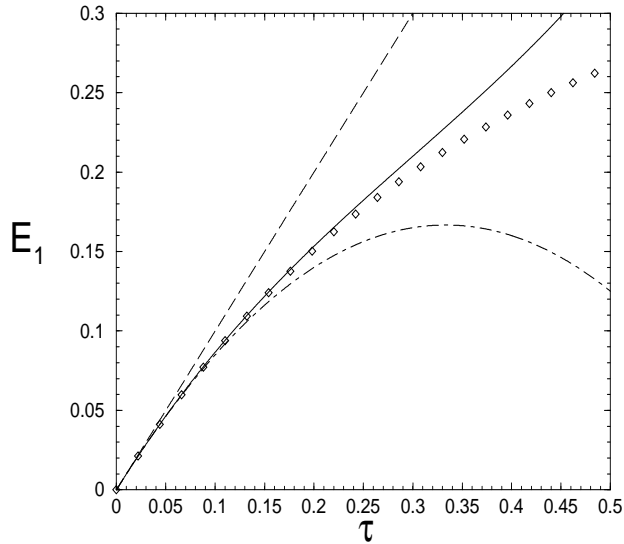,width=8cm,height=8cm}}
\vspace{1cm}
\caption{\label{fig2} 
Comparison of the approximations of $E_1$ 
obtained by 
truncating the series expansion (\ref{earser}) to 1st (dashed), 
2nd (dashed-dotted) and 3rd order (solid) with the exact 
solution 
(diamonds).
The relative feed rate is $h=1$.} 
\end{figure}

\begin{figure}[htbp]
\centerline{\psfig{figure=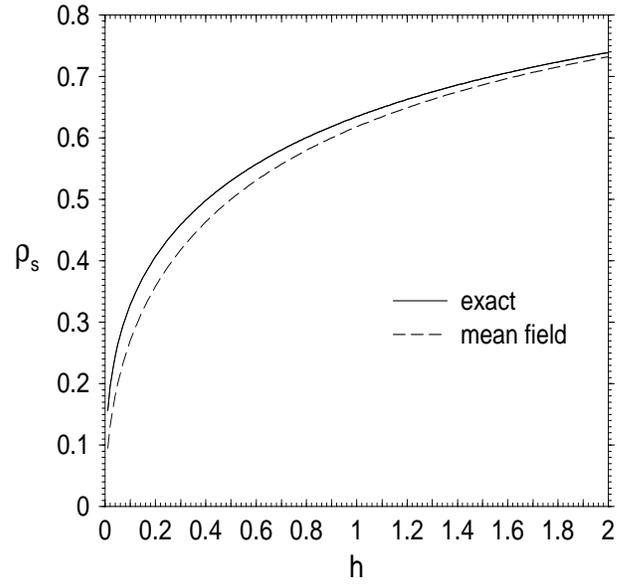,width=8cm,height=8cm}}
\caption{\label{fig3} 
Comparison of the steady state coverage
as a function of 
the relative feed rate $h$ computed from the exact 
solution and the MF 
approach.}
\end{figure}

The boundary condition at the origin determines the 
appropriate normalization factor 
for the steady state solution, 
$E_{n,s}$. The steady state coverage is then given by 

\be
\label{stbess}
\rho_s=1-\frac{J_{2h^{-1}+1}(2h^{-1})}{J_{2h^{-1}}(2h^{-1})}.
\ee
For 
$h\ll 1$ we can use the Taylor expansion of $J_{x+\Delta 
x}(x)$ around $x$ to 
obtain
\be
\rho_s\approx 
-\frac{1}{J_{2h^{-1}}(2h^{-1})} 
\left.\frac{d}{dx}J_x(2h^{-1})\right|_{x=2h^{-1}}.
\ee
We compute the derivative of the Bessel function by using 
the asymptotic form
\cite{Abra}
\be
\label{asbess}
J_\nu(\nu+z\nu^{1/3})=2^{1/3}\nu^{-1/3}\mbox{Ai}(-2^{1/3}z)+O(\nu^{-1}),
\quad 
\nu\to\infty, 
\ee
(where Ai($\cdot$) denotes the Airy 
function) and obtain an explicit expression for 
the small $h$ 
limit of the stationary coverage: 

\be
\label{stecon}
\rho_s=\frac{|\mbox{Ai}'(0)|}{Ai(0)}\,h^{1/3},\qquad 
h\to 0.
\ee 
In the opposite limit, $h\rightarrow \infty$, we have 
the following 
result:
\be
\label{largeh_ps}
\rho_s=1-h^{-1},\qquad h\to\infty.
\ee
Let us now examine the early 
time kinetics of our model. In general, it is difficult
to 
obtain a closed solution of the hierarchy (\ref{diffso}) 
because of the
nonconstant coefficient of $E_n$ on the rhs.  
Under the simplifying assumption of an
initially full 
lattice, we take the LT of (\ref{diffso}) and 
get
\be
\label{dyneq}
\hat{E}_{n-1}-(2+s+hn)\hat{E}_n+\hat{E}_{n+1}=0,
\ee
with 
$\hat{E}_0=1/s$ and $\hat{E}_\infty=0$. Exploiting again 
the analogy of 
Eqs.~(\ref{dyneq}) with the recursion 
relations for the Bessel functions and using
the boundary 
conditions leads to 

\be
\label{lapEn}
\hat{E}_n=\frac{1}{s}\,\frac{J_{(2+s)h^{-1}+n}(2h^{-1})}
{J_{(2+s)h^{-1}}(2h^{-1})}.
\ee
For 
$s\to 0$, one recovers (\ref{stbess}) by making use of the 
theorem 
$\lim_{s\to 0}s\,\hat{E}_n=E_n(\infty)$.  Since 
the conjugate time variable $s$
appears in the index of the 
Bessel functions, the exact inversion of (\ref{lapEn})
is quite involved. Instead, it is better to work with asymptotic 
formula. For $n=1$,
we can express $\hat{E}_1$ as a 
continued fraction \cite{Abra}:
\be
\label{lafrac}  
\hat{E}_1(s)=\left(\frac{1}{s}\right)\,
\frac{1}{s+2+h-\frac{\di 1}{\di s+2+2h-\frac{1}{\di 
s+2+3h-\ldots}}}. 
\ee
We can now 
expand the right-hand side in powers of $s^{-1}$ by using the Stieltjes 
method for 
J-fractions \cite{Wall} and find
\be  
\hat{E}_1(s)=\frac{1}{s^2}-\frac{2+h}{s^3}+
\frac{(2+h)^2+1}{s^4}+\O\left(\frac{1}{s^5}\right).
\ee
Inverting 
this expression term by term, we obtain the early time 
behavior of the 
empty-site coverage (c.f. Fig. 
\ref{fig2}):
\be
\label{earser}
\rho(\tau)=1-\tau+\left(1+\frac{h}{2}\right)\,\tau^2
-\frac{(2+h)^2+1}{6}\tau^3+\O(\tau^4),\qquad\tau\to 0.
\ee
The continued fraction representation of the LT is not suitable 
for obtaining the
long time asymptotics.  Instead, we make 
an exponential ansatz directly
in Eqs.~(\ref{diffso}):  

\be
\label{ltexp}
E_n(\tau)=E_{n,s}+a_n\,e^{-2\lambda\tau} 
.
\ee
The calculation of $\tau_R$ follows the guidelines 
laid by R\'acz for annihilation 
with input (see 
\cite{Racz} and paragraph \ref{singinp}). Substituting 
Eqs.~(\ref{ltexp}) 
in the set of differential-difference equations 
(\ref{diffso}), one 
finds that the 
coefficients $a_n$ obey 
the same set of equations as $E_n$, except 
that 
$1+nh/2$ is 
now replaced by $1-\lambda+nh/2$. 
Using the boundary 
condition at $n=0$, we find the equation for the 
possible 
values of $\lambda$: 
\be
J_{(1-\lambda)2h^{-1}} 
\left(2h^{-1}\right)=0.
\ee
All the solutions $\lambda_i(h)$ are 
positive, since the 
zeros $j_i^\nu$ of the Bessel functions 
satisfy the inequality 
$\nu<j_i^\nu$ \cite{Abra}, meaning that 
all homogeneous perturbations
around the steady state decay 
exponentially with time. The small $h$ 
limit is carried 
out using the asymptotic form
(\ref{asbess}). Setting 
$\nu=2(1-\lambda)h^{-1}$ and 
$z=\nu^{2/3}/(1-\lambda)$ yields 

\be
\mbox{Ai}(-2\,h^{-2/3}\lambda/(1-\lambda)^{1/3})=0.
\ee
 From 
this, a cubic equation for $\lambda$ can be 
extracted,
\be
\label{cubic}
\lambda_j^3+p\lambda_j-p=0;\qquad 
p=\frac{|a_j|^3}{8}h^2,
\ee
where $a_j$ are the zeros of the Airy 
function. Since the discriminant 
$\Delta=(p/3)^3+(p/2)^2$ 
is positive, this equation has two unphysical 
complex 
solutions and one real solution, the latter of which is given 
by 
Cardan's formula:
\be
\lambda_j=u+v; \qquad   
u=(p/2+\sqrt{\Delta})^{1/3},\quad 
v=-\frac{3p}{u}.
\ee
Expanding this expression for small $h$, we find the smallest 
eigenvalue
\be
\lambda_1=\frac{|a_1|}{2}\,h^{2/3},\qquad h\to 
0,
\ee
from which we 
obtain
\be
\label{invrt}
\tau_R^{-1}=2\,\lambda_1=|a_1|\, h^{2/3}+\O(h^{4/3}) .
\ee 
The 
corresponding MF equation for $\rho(\tau)$ may be 
obtained
following the procedure of the preceding subsection. We 
find
\be
\frac{d 
\rho}{d\tau}=-\rho^2+h\,(1-\rho).
\ee
Integrating and using the initial condition, 
we obtain  
\be
\label{tideco}
\rho^{MF}=\frac{\rho_+-\sigma\,\rho_-\, 
e^{-\eta\tau}}{1-\sigma\,
e^{-\eta\tau}},
\ee
where 
$\eta=\sqrt{h^2+4h}$, $\rho_\pm=(-h\pm\eta)/2$ and 
$\sigma=(\rho_0-\rho_+)/(\rho_0-\rho_-)$. 
The predicted long time coverage is 
$\rho^{MF}(\infty)\equiv  \rho_s^{MF}=\rho_+$. 
For small 
$h$, the steady state coverage is
\be
\rho_s^{MF}= 
h^{1/2}+\O(h),
\ee
whereas for large $h$ we get
\be 
\label{larghc}
\rho_s^{MF}=1-h^{-1}+\O(h^{-2}).
\ee
We now examine the 
transient behavior of the MF approximation.  Taking 
$\rho_0=1$, 
i.e., an initially full lattice, the early time 
kinetics is 
\be
\label{stbeh}
\rho^{MF}=1-\tau+\left(1+\frac{h}{2}\right)\tau^2+\O(\tau^3),\qquad\tau\to0,
\ee
while the long-time behavior for an arbitrary $\rho_0$ is found to 
be
\be
\rho^{MF}\approx \rho_{+}+\sigma\,\eta\,e^{-\eta 
\tau} ,
\qquad \tau\to\infty.
\ee
Thus, the inverse relaxation time is 
\be
\tau_R^{-1}\approx 2\,h^{1/2}, \qquad  
h\to 0, 
\ee
and 
\be
\tau_R^{-1}\approx h, \qquad  
h\to \infty.
\ee
It is worth making a few comments on the MF 
approximation. For $h=0$ (no input), the 
MF equation 
predicts in both cases a $t^{-1}$ decay towards an empty steady 
state.
However, if $h$ is finite, the decay becomes 
exponential with a relaxation time that
diverges as $h^{-1/2}$ for 
small $h$. This observation suggests that $h=0$ can 
be
viewed as a transition point, and one can try to account for 
spatial
fluctuations by some kind of phenomenological scaling 
assumption inspired by the
theory of critical phenomena 
\cite{Smol}. On the other hand, the limit $h\to 0$ for
fixed 
$R$ and $D$ considered here automatically realizes the 
continuum approximation
$a\to 0$. As expected, 
$\rho_s$ 
approaches zero in this limit, but the exact solution is larger 
than MF
($\rho_s\propto h^{1/3}$ vs. $\rho_s^{MF}\propto 
h^{1/2}$).  Moreover, the exact
solution remains larger than 
the MF approximation over the whole $h$ range ( 
Fig.
\ref{fig3}). This is not surprising, since the coalescence step 
responsible for
particle removal requires that  interacting 
particles occupy neighboring sites,
while in the MF 
approximation the effective range of the interaction is 
arbitrarily
large.  On the other hand, as $h$ becomes large, the exact 
steady state approaches
the MF curve\footnote{This can also 
be checked analytically by using the ascending
series 
expansion of the Bessel function 
$J_{\nu}(x)=\sum_{k=0}^{\infty}
\frac{(-x^2/4)^{k+\nu}}{k!\,\Gamma(\nu+k+1)}$} given by 
Eq.~(\ref{larghc}) (c.f.
Fig. \ref{fig3}). This result is 
quite natural, since the input mechanism involves
no spatial 
correlations. For sufficiently small $h$, the system relaxes 
more slowly
into the steady state than the MF prediction ( 
$\tau_R^{-1}\propto h^{2/3}$ as
opposed to 
$\tau_R^{-1}\propto h^{1/2}$ in the MF case). In the large $h$ limit, 
an
expansion of the real root of (\ref{cubic}) shows that 
$\tau_R^{-1}\propto h$, as
expected from the MF approximation.  
The results (\ref{stecon}) for the stationary
coverage and 
(\ref{invrt}) for the relaxation time in the small input 
limit are in
full agreement with the corresponding continuum 
approximation \cite{Doer}. However, one
expects a discrepancy for 
early times, as observed for reversible coalescence.

\section{Annihilation Models}
\label{annmod}

The basic 
model consists of diffusing point 
particles that annihilate 
upon contact. The particles hop with probability $2D/a^2$ 
to nearest neighbor sites; the only difference from simple 
coalescence is
that here {\it both} particles vanish 
instantaneously whenever they attempt to occupy
the same site.  
This reaction scheme conserves the parity of
the total 
number of particles. Parity changes in a given interval may only 
arise by
particle migration into or out of the interval.  
The evolution of $G_n(t)$, the
probability that $n$ 
consecutive sites contain an even number of particles, 
provides
detailed information about the annihilation process 
$A+A\rightarrow S+S$. Following
similar arguments as those for 
coalescence, one arrives at
\cite{Mass}     
\be
\label{differeq}
\frac{dG_n}{d \tau}= G_{n-1}-2\,G_n+G_{n+1}. 
\ee
The 
boundary conditions are $G_0=1$ and $0\le G_n \le 1$. Note 
the similarity
between the equations for $G_n$ (for 
annihilation) and $E_n$ (for coalescence). 
Indeed, it has been 
shown that the basic coalescence and annihilation models
lie in the same universality class \cite{Pel}; 
in the framework of our interval method, 
one can show that the only difference between
coalescence, annihilation and the zero-temperature
q-state Potts model stems from the specific form of the 
various initial conditions
\cite{Mass}. In the present case, 
if the lattice has a random distribution of
particles with 
global coverage $\rho_0$, a simple combinatorial argument 
yields
\be
\label{inGn}
G_n(0)=\frac{1}{2}+\frac{1}{2}(1-2\rho_0)^n.
\ee 

As predicted by early works of Torney and McConnell \cite{Torn}, and 
Lushnikov \cite{Lush}, the exact  solution of (\ref{differeq}) 
(on- and off-lattice) shows an 
anomalous, asymptotic
$\tau^{-1/2}$ decay of the lattice coverage $\rho=1-G_1$ to the 
empty steady state.  

The annihilation model may be 
extended by including  
input or birth processes. In the 
following, we present 
exact results
concerning some of these 
possibilities.

\subsection{Single particle 
input}
\label{singinp}

At each time step a particle is injected at a 
randomly 
chosen site at rate $Ra$. The site becomes occupied if it 
is initially empty 
and becomes empty otherwise. This is 
equivalent to adding the
toggle reactions $S\rightarrow A$ and 
$A\rightarrow S$ to the original
annihilation scheme.   The 
relevant set of equations for the probabilities 
$G_n$ now 
reads 

\be
\label{dynGn}
\frac{dG_n}{d\tau}=G_{n-1}-2\,G_n+G_{n+1}+n\,h(1-2G_n),
\ee
with the same boundary 
conditions as above. Here, the last term 
represents parity changes 
due to particle input, and
$\tau=2Dt/a^2$ and $h=Ra^3/2D$.

The model turns out to be closely related to the case of 
coalescence with input studied 
above.  Indeed, Eqs.~(\ref{dynGn}) 
may be solved with similar techniques to those 
used in 
\ref{coainp}.  The stationary coverage can be expressed in 
terms 
of Bessel functions, and for low $h$ it is smaller than 
(\ref{stecon}) by a 
factor of $2^{-2/3}$.  The relaxation 
time turns out to be smaller 
than that for coalescence by 
a factor of $2^{-2/3}$ \cite{Doer}.  

These results are in 
full agreement with previous calculations by R\'acz, who 
provided 
an exact solution by mapping the above dynamics into 
a kinetic Ising model
\cite{Racz}.  In R\'acz's model, the time evolution of an infinite 1D 
array $\{\sigma\}\equiv \{\ldots,\sigma_i,\sigma_{i+1},\ldots\}$ of
stochastic pseudospin variables $\sigma_i(t)=\pm 1$ is considered; 
particles are identified with domain 
walls between regions containing up or down spins only, i.e., the
bond variables $n_i\equiv (1-\sigma_i\sigma_{i+1})/2$ are seen as 
particle occupation numbers. For random homogeneous initial conditions,
translational invariance holds, and the mean (global) particle coverage 
is given by $\rho(t)=(1-\langle \sigma_i \sigma_{i+1}\rangle)/2$. The 
state probability $P(\{\sigma\},t)$ satisfies the Glauber master 
equation
\be
\label{meqsp}
\frac{dP(\{\sigma\},t)}{dt}=\sum_{i=-\infty}^{\infty}\sum_{\alpha=1}^2
[w_i^{(\alpha)}(\{\sigma\}_i^\alpha)\,P(\{\sigma\}_i^\alpha,t)-
w_i^{(\alpha)}(\{\sigma\})P(\{\sigma\},t)],
\ee
where the state $\{\sigma\}_i^1$ is obtained from $\{\sigma\}$ by 
flipping the $i$th spin, and $\{\sigma\}_i^2$ differs from $\{\sigma\}$
by the simultaneous flipping of all spins $\sigma_j$ with $j\le i$. The 
flipping rates are given by 
\be
w_i^{(1)}(\{\sigma\})=\frac{\Gamma}{2}\, [1-\frac{\delta}{2}\, 
\sigma_i (\sigma_{i+1}
+\sigma_{i-1})], \qquad w_i^{(2)}(\{\sigma\})=\Gamma h.
\ee
Setting $\delta=1$ and $\Gamma=2D/a^2$ corresponds exactly to the 
dynamics of our model with a relative feed rate $h$. 

It has also been suggested that annihilation with single 
particle input may be
relevant to the kinetics of a  
certain class of processes involving cluster-cluster
aggregation 
in the presence of sources and sinks. For example, in 
aerosol formation
aggregation centers can be generated by 
photo-oxidation, and large clusters may
disappear as a result of 
sedimentation \cite{Fried}. According to R\'acz, it 
should be expected that such models be in the same universality class 
as annihilation with
input
\cite{Racz}.

\subsection{ 
Input of adjacent pairs }

Pairs of particles are injected 
simultaneously at adjacent 
sites at rate $R$ per site per unit 
time. Thus, the steps $SS\rightarrow AA$,
$AS\rightarrow 
SA, SA\rightarrow AS$ and $AA\rightarrow SS$ take place at 
equal
rates (but generically different from the hopping rate 
for diffusion and  
annihilation events).  The kinetics is 
described 
by the equations 

\be
\label{differeq2}
\frac{dG_n}{d \tau}= G_{n-1}-2\,G_n+G_{n+1}+2h(1-2G_n), 
\ee  
with $\tau=2Dt/a^2$ and $h=Ra^2/(2D)$. The boundary 
conditions remain the
same as before.  

Again, we begin our study 
of this model with the steady state solution:  

\be
G_{n,s}=\frac{1}{2} + \frac{1}{2}\left[ 1+2h-\sqrt{(1+2h)^2-1} 
\right]^n.
\ee
This implies a steady state coverage of 

\be
\label{stcov}
\rho_s=\sqrt{h^2+h}-h.
\ee
For fast input 
rates (large $h$), we have 

\be
\label{lahcov}
\rho_s\approx\frac{1}{2}-\frac{1}{8h},\qquad h\to\infty,
\ee
while for small $h$, 

\be
\rho_s\approx h^{-1/2},\qquad h\to0. 
\ee 
We now 
derive the exact time dependence of the concentration for a 
random homogeneous initial distribution. Applying the LT to 
both sides of (\ref{differeq2}) and using the initial 
condition (\ref{inGn}) yields a set of inhomogeneous difference 
equations. 
However, the inhomogeneity does not depend on $n$ 
and can be easily shifted away. 
After using the boundary 
conditions, we obtain 

\be
\label{LapG}
\hat{G}_n=\frac{1}{2s}+\frac{(1-2\rho_0)}{2(s+a-b)}+
\left(\frac{1}{2s}-\frac{1}{2(s+a-b)}\right)\lambda_-^n, 
\ee 
where $a=2(1+2h)$, $b=1-2\rho_0 + 1/(1-2\rho_0)$, and 

\be
\lambda_-=\frac{s+a-\sqrt{(s+a)^2-4}}{2}.
\ee
To 
obtain the explicit time dependence of the probabilities 
$G_n$, we invert Eqs. 
(\ref{LapG}) using the convolution 
theorem for the LT. We thus get    

\be
\label{explGn}
G_n(\tau)=\frac{1}{2}+\frac{n}{2}
\int_0^\tau 
\frac{e^{-a\tau'}\,I_n(2\tau')}{\tau'}\,d\tau'+
\left[\frac{(1-2\rho_0)^n}{2}\,-\frac{n}{2}
\int_0^\tau 
\frac{e^{-b\tau'}\,I_n(2\tau')}{\tau'}\,d\tau'
\right]e^{(b-a)\tau}.
\ee  

In the limit $\rho_0\to 1/2$, the difference equations for 
the $\hat{G}_n$
become homogeneous; the last term 
in~(\ref{explGn}) vanishes and one 
need compute only one remaining 
integral.  

Determining the long-time asymptotics of the 
solution (\ref{explGn}) requires 
separate analysis of two 
cases:  $\rho_0 \leq 1/2$ and $\rho_0 > 1/2$. 
As we shall 
see, both cases yield the same expansion for 
$\rho(\tau)$.

For $n=1$, the integrals in the rhs of (\ref{explGn}) can 
be computed by making use 
of the identity 
\cite{Luke}
\be
\label{intbes}
\int_0^z \frac{e^{-\beta y}I_1(2 
y)}{y}\,dy=e^{-\beta z} 
\sum_{k=0}^\infty \frac{z^{2k+1}}
{k! 
(2k+1)\Gamma(k+2)} \left.\right._1\!\!\,F_1(1;2(k+1);\beta 
z),
\ee 
where $\left.\right._1\!\!\,F_1(\alpha;\beta;\gamma)$ is 
a generalized 
hypergeometric function. For short times 
$\tau$, however, it is easier to use
the series expansion of 
the Bessel functions to evaluate the integrands. 
In this 
limit, we obtain, to second order: 

\be
\label{stexp}
\rho(\tau)=\rho_0-2(\rho_0^2+h(2\rho_0-1))\tau-(2\rho_0^3+
\rho_0^2(1+8h)+8h^2\rho_0-4h^2-h)\tau^2+\O(\tau^3).
\ee
The ascending series on the rhs of (\ref{intbes}) is not 
suitable for the analysis 
of the long-time asymptotics. Instead, 
we observe that for
$\rho_0\le\frac{1}{2}\,\,(b\ge 2)$,   
the integrals in (\ref{explGn}) converge 
for
$\tau\to\infty$.  We can  therefore split them into a definite and an 
indefinite part and use the adequate  asymptotic expansion for 
the latter. This 
yields
\be
\label{ascov}
\rho(\tau\to\infty)=\rho_s+(16\sqrt{\pi})^{-1}\,\left[\frac{1}{h}
-\frac{1-2\rho_0}{\rho_0^2}\right]\tau^{-3/2}\,e^{-4h\tau}, 
\qquad 
\rho_0\le\frac{1}{2}.
\ee 
For 
$\rho_0>\frac{1}{2}\,$ $(b<-2)$, however, the last integral diverges. 
Nevertheless, we can split the integral as 
follows:
\be
\int_0^\tau 
\frac{e^{-b\tau'}\,I_1(2\tau')}{\tau'}\,d\tau'=
\int_0^{C_1} 
\frac{e^{-b\tau'}\,I_1(2\tau')}{\tau'}\,d\tau'+
\int_{C_1}^\tau 
\frac{e^{-b\tau'}\,I_1(2\tau')}{\tau'}\,d\tau',
\ee
with a sufficiently large constant $C_1$, so that the 
second integral 
on the rhs can be well approximated by 
expanding the modified Bessel 
function to its leading term 
\cite{Abra}. We can then replace 
the original integral by  

\be
\frac{1}{2\sqrt{\pi}}\int_{C_1}^{\tau} 
\frac{e^{(2-b)\tau'}}{\tau'^{3/2}}  
\approx 
C_2-\frac{1}{\sqrt{\pi\tau}}e^{(2-b)\tau}+
\sqrt{2-b}\,\,\mbox{erfi}(\sqrt{(2-b)\tau}),
\ee
where 
$C_2$ is a constant and the imaginary error function is 
defined 
as 

\be
\mbox{erfi}(x)=-i\,\mbox{erf}(ix)=\frac{2}{\sqrt{\pi}}\int_0^x e^{y^2}\,dy.
\ee
Thus, for $n=1$, the 
last term in (\ref{explGn}) becomes 

\be
\label{secexp}
-\frac{e^{(b-a)\tau}}{2}
\int_0^\tau\frac{e^{-b\tau'}\,I_1(2\tau')}{\tau'}\,d\tau' 
\approx
-\sqrt{\frac{2-b}{\pi}}\,D(\sqrt{(2-b)\tau})\,e^{(2-a)\tau}
+\frac{1}{2\sqrt{\pi\tau}}\,e^{(2-a)\tau}
+C_3\,e^{(b-a)\tau},
\ee
where 
$C_3=-C_2/2$ and $D(x)=e^{-x^2}\int_0^x e^{y^2}\,dy$ is 
the
so-called Dawson's integral \cite{Abra}. For large 
arguments,  

\be
\label{asdaw}
D(x)=\frac{1}{2x}+\frac{1}{4x^3}+\O\left(\frac{1}{x^5}\right). 
\ee
If we now use 
(\ref{asdaw}) to expand the first term on the rhs of 
(\ref{secexp}), 
we end up with 
\be
-\frac{e^{(b-a)\tau}}{2}\int_0^\tau 
\frac{e^{-b\tau'}\,I_1(2\tau')}{\tau'}\,d\tau'
\approx 
-\frac{1}{4(2-b)\tau^{3/2}}\,e^{(2-a)\tau}+
\O(\tau^{-5/2}\,e^{(2-a)\tau})+C_3\,e^{(b-a)\tau}.
\ee
Clearly, 
the last, purely exponential term decays faster than all 
other terms, 
since $b<-2$. Thus, the leading order is the 
term proportional to $\tau^{-3/2}$.  We then find the same 
decay law
as for $\rho_0<1/2$, of Eq.~(\ref{ascov}). The relaxation 
time is universal and given by $\tau_R=1/(4h)$.

The MF approximation is derived by noting that 
$G_1=1-P_1$ and 
$G_2=1-2\,P_1+2\,P_2$ and using the factorization 
ansatz 
$P_2=P_1^2$. This leads to: 

\be
\frac{d\rho}{d\tau}=-2\rho^2+2h\,(1-2\rho)
\ee
The physical steady state of this equation is given by (\ref{stcov});
this is not surprising, since the steady state is an equilibrium one,
like in the reversible coalescence case. The early time behavior is 
in agreement with Eq.~(\ref{stexp}) up to the first order, while 
the long-time approach of the MF solution to $\rho_s$ is easily 
found to be  
\be
\rho^{MF}(\tau)-\rho_s
\propto 
2\frac{(h^2+h)^{1/2}(\rho_0-\rho_s)}
{\rho_0+(h^2+h)^{1/2}+h}\,e^{-4(h^2+h)^{1/2}\tau},\qquad\tau\to\infty,
\ee
Once again, the MF approach fails to capture the time dependent
prefactor of the exponential in the exact solution, although it 
yields the correct relaxation time for large~$h$.

In the small $h$ limit, the expressions for $\rho_s\approx h^{1/2}$ and 
$\tau_R$ agree with the continuum limit \cite{Mass2}. These results have
also been obtained by R\'acz by mapping the above model into the 
kinetic Ising model; to do so, one chooses $\Gamma=2D/a^2$, 
$\delta=(1-h)/(1+h)$, and $w_i^{(2)}(\{\sigma\})=0$  in 
Eq. (\ref{meqsp}) \cite{Racz}. This yields hopping to nearest sites at
rate $D/a^2$, nearest-neighbor annihilation at  rate $2D/(a^2 (1+h))$,
and pair production at rate $2Dh/(a^2 (1+h))$, i.e., the dynamics of 
our model for small $h$. At larger $h$, however, the correspondence 
is lost because annihilation events in the spin model are slower 
than hopping onto empty sites. Our results are a natural 
generalization of R\'acz's solution for the case of arbitrary $h$. 
In particular, in the large $h$ limit, the steady state coverage 
is given by Eq.~(\ref{lahcov}).

We also find that the decay to the steady state 
is not purely exponential but rather given by a power of 
$\tau$ times an exponential. This corroborates a recent 
conjecture of Habib {\it et al.} \cite{Habib}, based on 
the continuum approximation of Eq.~(\ref{differeq2}).  

Finally, 
we remark that the above model might capture essential 
features of transient optical absorption at energies less 
than the interband gap in certain 1D organic semiconductors  
\cite{Su,Orens}. It has been argued that the high 
energy 
peak in the absorption spectrum of {\it trans}-polyacetylene 
may be due to the 
photogeneration of intrinsic, 
self-localized excitations
of the semiconductor chain \cite{Orens} in 
form of alternating soliton-antisoliton 
pairs which move 
randomly along the chain under the influence of thermal 
fluctuations and annihilate upon contact.

\subsection{Annihilation with Symmetric Birth}
The input process consists 
of a double particle birth, 
i.e., a particle produces 
offspring at both neighboring sites at rate V (on each
side). 
In other words, we have the additional reactions 
$SAS\rightarrow AAA$, 
$SAA\rightarrow AAS, AAS\rightarrow SAA$ and 
$AAA\rightarrow SAS$.  The probabilities
$G_{n}(t)$ obey the 
equations \cite{Mass2}:
\bml
\label{setGn}
\be
\label{dynGn2}
\frac{dG_{n}}{dt}=(2D+2V)(G_{n+1}-2G_{n}+G_{n-1}), 
\qquad n>1.                
\ee    
The boundary condition is 
nontrivial; for $n=1$, one 
has
\be
\label{bcGn}
\frac{dG_1}{dt}=2D\,(G_{2}-2G_{1}+1) + 
2V(G_{2}-G_{1}).
\ee
\eml
Notice that in the above equations we have set $a=1$. This is 
done bearing in mind that there is no 
straightforward continuum 
approximation of 
the model, due to 
the form of the boundary condition.  As before, we 
have the 
condition, $0\leq G_{n} \leq 1$. Again, we solve the set 
of Eqs.~(\ref{setGn}) for an initially uncorrelated 
distribution 
given by the initial condition~(\ref{inGn}). 

We study this model with and 
without 
diffusion.  
Let us first consider the case of $D=0$.  
In this limit, clusters 
of particles 
can spread only by 
giving birth. To simplify the evolution equations 
for 
the
$G_n$ we let $\tau=2Vt$ and 
$G_{n}(\tau)=\frac{1}{2}(1+F_{n}(\tau))$.  
We then have the initial 
BVP:
\bml
\ba
\frac{dF_n}{d\tau}&=&F_{n+1}-2F_{n}+F_{n-1},\\
\frac{dF_1}{d\tau}&=&F_{2}-F_{1},\\
F_{n}(0)&=&(1-2\rho_0)^n.
\ea
\eml
The solution of these equations is readily obtained by 
LT methods similar to those used above. The LT of $F_1(t)$
takes the form 

\be
\hat{F}_{1}(s)=\frac{1-2\rho_0}{s-\upsilon}-\frac{\rho_0}{s-\upsilon}
\left(\sqrt{\frac{s+4}{s}}-1\right).
\ee 
where $\upsilon\equiv 4\rho_0^2/(1-2\rho_0)$. For three simple cases, 
$\rho_0=0,$ $\rho_0=1/2,$
and $\rho_0=1$, 
the transform can be inverted to yield 
$G_{1}=1$, $G_{1}=1/2$,  
and 
$G_{1}=1/2\,(1-e^{-2\tau}I_{0}(2\tau))$, respectively.  These results 
merely indicate 
that an empty system remains empty; $\rho_0=1/2$ is the 
steady 
state coverage; and initially full lattices decay into 
the steady 
state.

We can also invert explicitly 
$\hat{F}_1(s)$ when $0<\rho_0<1/2$.  
We get
\be
F_{1}(\tau)=e^{\upsilon\tau}\left[ 
1-\rho_0-\rho_0(\upsilon+4)\int_{0}^{\tau}
e^{-(\upsilon+2)\tau'}I_{0}(2\tau')d\tau' \right]-\rho_0 
e^{-2\tau}I_{0}(2\tau).
\ee
Using the appropriate long time expansion, we get
\be
F_{1}(\tau)\approx\frac{1-2\rho_0}{\rho_0\sqrt{4\pi\tau}},\qquad
  \tau\to\infty,
\ee
or, in terms of the 
coverage:
\be
\label{ascov2}
\rho(\tau)=\frac{1}{2}-\frac{1}{2}F_1(\tau)
\approx\frac{1}{2}-\frac{1-2\rho_0}{4\rho_0\sqrt{\pi\tau}},\qquad
\tau\to\infty.
\ee
Thus the system decays algebraically 
into the steady state 
$\rho_s=1/2$.  In fact,
Sudbury had 
shown that any homogeneous, random distribution leads to a 
half-empty lattice \cite{Sudb,Mass2}. It is therefore not 
surprising that 
we obtain the same 
behavior also for 
$1/2<\rho_0<1$.  In this regime, $\upsilon<0$, and 
the LT 
inversion becomes difficult.  Since we are mainly interested 
in 
the long time asymptotics of $F_{1}$, we circumvent this 
difficulty 
by expanding $\hat{F}_{1}(s)$ about $s=0$ and 
making use of a 
Tauberian theorem. For $s\rightarrow 0$, 

\be
\hat{F}_{1}(s)\approx 
\frac{2\rho_0}{\upsilon\sqrt{s}}+\O(1),
\ee
implying that 
\be
F_{1}(\tau) \approx 
\frac{1-2\rho_0}{\rho_0\sqrt{4\pi\tau}},\qquad\tau\to\infty.
\ee
This matches the result for 
$0<\rho_0<1/2$. 

We now 
turn to the case of annihilation with double birth and 
diffusion, $D>0$.
Again, we take  Eqs.~(\ref{setGn}) and 
(\ref{inGn}) as a starting point; setting $\tau = 
2(D+V)t$ and $G_{n}=\frac{1}{2}(F_{n}+1)$, and 
taking the LT with respect to $\tau$, we finally obtain

\be
\hat{F}_{1}(s)=\frac{1}{2}\left(\frac{\kappa}{s}-\frac{2\rho_0(1-\kappa)+
\kappa}{s-\upsilon}\right) 
\left(\frac{\sqrt{s^2+4s}-s-2\kappa}{(1-\kappa)s-\kappa^2}\right)+
\frac{1-2\rho_0}{s-\upsilon}.
\ee 
with $\kappa \equiv D/(D+V)$. Notice that we recover the 
appropriate solutions when 
$\kappa=1$ and $\kappa=0$.  
These limits refer to the $V=0$ and $D=0$ 
cases, respectively.  
$\hat{F}_{1}(s)$ can be inverted for 
particular cases of the parameters 
$\rho_0$ and $\kappa$. 
However, it is simpler to make use of the $s=0$ expansion to 
obtain the long time asymptotics:  

\be
\hat{F}_{1}(s)\approx\frac{1}{s}-\frac{1}{\kappa\sqrt{s}},\qquad s\to0. 
\ee
This leads to 

\be
\label{covlt}
\rho\approx\frac{1}{2\kappa\sqrt{\pi\tau}},\qquad\tau\to\infty.
\ee
This result is not 
valid for $\kappa=0$, since
the limit $\kappa\rightarrow 0$ is singular. 
It is easy to see that the 
system
cannot become empty 
without diffusion \cite{Sudb,Avr3}. In that case, (\ref{ascov2}) 
holds.  
 
Let us compare the above results with the MF 
approximation, which can be 
derived from the equation for 
$G_1$ in the usual way. In the presence of diffusion, 
the simple MF approach yields
\be
\label{mfdiff}
\frac{d\rho}{d\tau}=(1-\kappa)\rho-2\rho^2,
\ee
 Consequently, the steady states are 
$\rho_s=0, (1-\kappa)/2$. Note that the existence of a non-empty 
steady state is already in contradiction to the exact 
solution (\ref{covlt}). The nontrivial, time dependent 
solution of (\ref{mfdiff}) reads 
\be
\rho^{MF}=\frac{(1-\kappa)\,\rho_0}{2\rho_0+(1-\kappa-2\rho_0)\,
e^{(\kappa-1)\tau}},
\ee
predicting exponential relaxation, in contrast to the inverse power 
law behavior of the exact solution. Thus, for $\kappa>0$, the MF 
approximation fails both at the static and the dynamic levels,
whereas for $\kappa=0$ (no diffusion), it describes the 
steady state correctly, but not the long time dynamics.  
At the MF level, the effect of diffusion consists in driving 
the system from a steady state
with half the lattice filled into a steady 
state with a smaller (but finite!) coverage. In contrast, 
the effect of diffusion on the exact solution is
much more drastic, since the slightest amount of 
mobility already lands 
the system in an empty absorbing state. Note also the 
somewhat surprising fact that the presence of diffusion 
drives the steady state away from the 
MF prediction, except when the value of $\kappa$ becomes close 
to one. In this limit, $\rho^{MF}(\infty)$ again becomes arbitrarily 
close to zero.

\section{Summary and outlook}
\label{sumout}

We have derived a 
series of exact results for the 1D kinetics of the lattice 
coverage and the particle concentration for coalescence 
and annihilation with various particle sources. 

In the 
case of reversible coalescence the MF approach 
reproduces 
the equilibrium steady state correctly, but it only yields 
the 
 relaxation time of the exact solution if $h$ and 
$\rho_0$ are large enough. 
Besides, this approximation is not 
sensitive enough to reproduce the observed 
fluctuation-induced phase transition. In contrast, the off-lattice 
approximation 
valid for low input rates $h$ (and thus for low 
transient concentrations) 
provides a good 
qualitative 
description of the long time asymptotics. For large $h$, no 
qualitative changes in the $c_0$ and $\tau$ dependence are 
observed, but the values 
of the prefactor and the relaxation 
time in the asymptotic form of the solution
depend on the 
lattice constant $a$.  
For early times, we find that the 
exact solution is in agreement with the 
classical MF 
approach, while the off-lattice approximation dictates a 
slower 
kinetics in this regime (one expects a similar conclusion 
for
the other lattice reactions studied in this paper).

In 
the case of coalescence and annihilation with single-particle 
input, 
the exact steady state is expressible in terms of 
Bessel functions. For small 
$h$-values, it is well 
reproduced by the continuum approach, 
whereas for large values, it 
becomes MF-like. In this case, the relaxation is 
purely 
exponential, and $\tau_R$ is again MF-like only for 
large 
$h$. 

For annihilation with pair input, the steady state is 
an 
equilibrium state and can be derived directly from the 
MF approach. 
However, the relaxation is not purely 
exponential, as predicted by the MF approach, since the 
concentration 
behaves like 
$\tau^{-3/2} e^{-\tau/\tau_R}$ with a 
universal $\tau_R$. In this case, 
the somewhat faster 
relaxation with respect to the single particle case 
may be due 
to the higher particle input. 

Summarizing, in the cases 
for which the continuum approximation was considered, 
the 
latter is in good qualitative agreement with 
the exact long 
time kinetics and with the steady state in the low $h$ 
limit, 
however, deviations from the exact kinetics are 
expected for large 
values of $h$, i.e., at higher transient 
concentrations. In this limit, the MF approximation yields 
relatively 
good results for $\tau_R$; even in the reversible 
coalescence 
case, provided that the initial concentration 
is sufficiently high. 
Fluctuations can also be safely 
neglected in the short time regime. 

As for annihilation with 
symmetric birth, we were able to confirm 
Sudbury's 
result for an infinite lattice, and prove that the long 
time relaxation to the steady 
state is proportional to 
$\tau^{-1/2}$, both for immobile reactants and 
in the presence of 
diffusion. In the diffusionless case, the MF 
approach
reproduces the exact steady state correctly, but not the long 
time kinetics. 
Any amount of diffusion lands the system in an 
empty state, 
in contrast to the MF prediction.   

Possible generalizations of our work include studying finite size 
effects 
and generalizing the method to better understand 
the spatial organization 
in these systems; in this respect, 
it is of interest to compute quantities such as 
interparticle distribution functions or multiple-point 
correlations 
from joint probabilities for pairs of intervals, both 
on-lattice
and in the continuum limit \cite{Lin}. The present methods also 
allow to study the effect of temporal
correlations \cite{Gryn} or 
inhomogeneities in the initial conditions \cite{Ali}. 
 
\acknowledgements

We are grateful to G. Nicolis for helpful suggestions. E.A. was
supported in part by the Interuniversity Attraction Poles program
of the Belgian Federal Government. D.b-A. gratefully acknowledges 
the NSF for support, under contract no.\ PHY-9820569.  

\newpage

\newpage

\begin{thebibliography}{99}

\bibitem{Priv0} V. Privman (Ed.) {\it Nonequilibrium Statistical 
Mechanics in One Dimension}, Cambridge University Press, Cambridge
(1996).

\bibitem{Schu} G. M. Sch\"utz in {\it Phase Transitions and Critical 
Phenomena}, vol. {\bf 19}, C. Domb and J. Lebowitz (eds.), Academic
Press, London (2000). 

\bibitem{Marr} J. Marro and R. Dickman, {\it Nonequilibrium Phase 
Transitions in Lattice Models}, Cambridge University Press, Cambridge
(1999).

\bibitem{Hen1} M. Henkel, E. Orlandini, and J. Santos, Ann. Phys. 
{\bf 259}, 163 (1997).

\bibitem{Alc} F.C. Alcaraz, M. Droz, M. Henkel, and V. Rittenberg, 
Ann. Phys. {\bf 230}, 250 (1994).

\bibitem{Cadi} V. Privman, A.M. Cadilhe, and M.L. Glasser, J. Stat. 
Phys. {\bf 81}, 881, (1995).

\bibitem{Avr} 
D. ben-Avraham, M. Burschka and C. Doering,
J. Stat. Phys. {\bf 60}, 695 (1990). 

\bibitem{Avr2} M. Burschka, C. Doering and D. ben-Avraham, 
Phys. Rev. Lett. {\bf 63}, 700 (1989).

\bibitem{Doer} C.R. Doering and D. 
ben-Avraham, Phys. Rev. Lett. {\bf 
62}, 2563 (1989)

\bibitem{Kreb} K. Krebs, M.P. Pfannm\"uller, B. Wehefritz, and H.
Hinrichsen, J. Stat. Phys. {\bf 78}, 1429 (1995).

\bibitem{Sim} H. Simon, J. Phys. {\bf A28}, 6585 (1995).  

\bibitem{Mobi} M. Mobilia and P.A. Bares, cond-mat/0107427.

\bibitem{Torn} D. C. Torney and H. M. McConnell, J. Phys. 
Chem. {\bf 87}, 1941 (1983).

\bibitem{Lush} A.A. Lushnikov, 
Phys. Lett. {\bf A120}, 135 (1987).

\bibitem{Racz} Z. R\'acz, Phys. Rev. Lett. {\bf 55}, 1707 
(1985).

\bibitem{Spou} J.L. Spouge, Phys. Rev. Lett. {\bf 60} 871 
(1988).

\bibitem{Fam} F. Family and J.G. Amar, J. Stat. Phys. {\bf 65}, 1235 
(1991).

\bibitem{Redn} S. Redner and K. Kang, Phys. Rev. {\bf 
A32}, 435 (1985).

\bibitem{Priv} V. Privman, Phys. Rev. {\bf E50}, 50 
(1994).

\bibitem{Gryn} M.D. Grynberg and R.B. 
Stinchcombe, Phys. Rev. Lett. {\bf 76}, 851 (1996).

\bibitem{Hen2} M. Henkel and H. Hinrichsen, J. Phys. {\bf A34},
1561 (2001). 

\bibitem{Lin} J.C. Lin, C.R. Doering and D. ben-Avraham, Chem. Phys. 
{\bf 146}, 355 (1990).

\bibitem{Pes} I. Peschel, V. Rittenberg, and U. Schultze, 
Nucl. Phys. {\bf B430}, 633 (1994). 

\bibitem{Ali} M. Alimohammadi, M. Khorrami, A. Aghamohammadi, 
cond-mat/0105124.

\bibitem{Mass} T. Masser and D. ben-Avraham,  
Phys. Lett. {\bf A275}, 382 (2000).

\bibitem{Mass2} T. Masser and D. 
ben-Avraham, Phys. Rev. {\bf E63}, 066108 
(2001).

\bibitem{Habib} 
S. Habib, K. Lindenberg, G. Lythe and C. Molina-Paris, 
preprint cond-mat/0102270.  

\bibitem{Abad} E. Abad, H. L. Frisch and G. Nicolis, J. Stat. Phys. 
{\bf 99}, 1397 (2000).

\bibitem{Smol} Z. R\'acz, Phys. Rev. Lett. {\bf A32}, 1129 (1985).

\bibitem{Sudb} A. Sudbury, Ann. Prob. {\bf 18}, 581 (1990).

\bibitem{Lin2} J.C. Lin, Phys. Rev. {\bf A45}, 3892 
(1992).

\bibitem{Cars} H. S. Carslaw and J. C. Jaeger, {\it Conduction of 
Heat in Solids}, 2nd. Ed., Clarendon, Oxford (1959).

\bibitem{Doe} G. Doetsch, {\it Theorie und Anwendung der 
Laplace Transformation}, Dover Pub., New York (1943).

\bibitem{Fell} W. Feller {\it An introduction to Probability Theory 
and its Applications},
vol. {\bf 2}, 2nd Ed., Wiley, New York (1971).

\bibitem{Abra} M. Abramowitz and I. Stegun, {\it 
Handbook of mathematical functions}, Dover pub., New York 
(1972).

\bibitem{Wall} H.S. Wall, {\it Analytic theory of 
continued fractions}, D. van Nostrand Comp. (1948).

\bibitem{Pel} L. Peliti, J. Phys. A {\bf 
19}, L365 (1986).

\bibitem{Fried} 
K. Friedlander, {\it Smoke, Dust and Haze: Fundamentals 
of Aerosol Behavior}, Wiley, New York (1977).

\bibitem{Luke} Y.L. Luke, {\it Integrals of Bessel functions}, 
Mac-Graw Hill, New York (1962). 

\bibitem{Su} W.P. Su and J.R. 
Schrieffer, Proc. Natl. Acad. Sci. U.S.A. 
{\bf 77}, 5626 (1980).

\bibitem{Orens} J. Orenstein and G. L. Baker, Phys. 
Rev. Lett. {\bf 
49}, 1043 (1982).   

\bibitem{Avr3} D. 
ben-Avraham, F. Leyvraz, and S. Redner, Phys. Rev. 
{\bf E50}, 1843 (1994).    

\end{thebibliography}
\end{document}